\documentclass[graybox]{svmult}

\usepackage{type1cm}        
\usepackage{makeidx}         
\usepackage{graphicx}        
\usepackage{multicol}        
\usepackage[bottom]{footmisc}

\usepackage{newtxtext}       %
\usepackage[varvw]{newtxmath}       

\makeindex             

\newcommand{\dd}{\textup{d}}
\def\eps{\varepsilon}
\def\E{\mathbb{E}}
\def\P{\mathbb{P}}
\def\R{\mathbb{R}}
\def\target{\textup{target}}
\def\oned{\textup{1d}}

\def\threed{\textup{3d}}
\def\PP{\mathcal{P}}
\def\dmin{d_{\textup{min}}}

\begin{document}

\title*{Competition of many searchers}
\author{Sean D. Lawley}
\institute{Sean D. Lawley \at Department of Mathematics, University of Utah, Salt Lake City, UT 84112 USA, \email{sean.lawley@utah.edu}}
%
%
\maketitle

\abstract*{First passage times (FPTs) are often used to study timescales in physical, chemical, and biological processes. FPTs generically describe the time it takes a random ``searcher'' to find a ``target.'' In many systems, the important timescale is not the time it takes a single searcher to find a target, but rather the time it takes the fastest searcher out of many searchers to find a target. Such fastest FPTs or extreme FPTs result from many searchers competing to find the target and differ markedly from FPTs of single searchers. In this chapter, we review recent results on fastest FPTs. We show how fastest FPTs depend on the mode of stochastic search (including search by diffusion, subdiffusion, superdiffusion, and discrete jumps), the initial searcher distribution, and properties of the spatial domain.}

\abstract{First passage times (FPTs) are often used to study timescales in physical, chemical, and biological processes. FPTs generically describe the time it takes a random ``searcher'' to find a ``target.'' In many systems, the important timescale is not the time it takes a single searcher to find a target, but rather the time it takes the fastest searcher out of many searchers to find a target. Such fastest FPTs or extreme FPTs result from many searchers competing to find the target and differ markedly from FPTs of single searchers. In this chapter, we review recent results on fastest FPTs. We show how fastest FPTs depend on the mode of stochastic search (including search by diffusion, subdiffusion, superdiffusion, and discrete jumps), the initial searcher distribution, and properties of the spatial domain.}

\section{Introduction}

Many events in physical, chemical, and biological systems are initiated when a ``searcher'' finds a ``target,'' which is called a first passage time (FPT) \cite{redner2001}. Depending on the system, the searcher could be, for instance, an ion, protein, cell, or predator, and the target could be a receptor, ligand, cell, or prey.  The majority of prior work in this area studies the FPT of a given single searcher. However, the relevant timescale in many systems is the time it takes the fastest searcher out of many searchers to find a target, which is called a fastest or extreme FPT \cite{weiss1983, yuste1996, yuste2000, yuste2001, redner2014, meerson2015, ro2017, basnayake2019, schuss2019, coombs2019, redner2019, sokolov2019, rusakov2019, martyushev2019, tamm2019, lawley2020esp1, lawley2020uni, lawley2020dist, morgan2023}.

Concretely, let $\tau$ denote the FPT to a target $U_{\target}$ of a single searcher whose stochastic position at time $t\ge0$ is denoted by $X(t)$. Mathematically, 
\begin{align}\label{eq:tau0}
    \tau
    :=\inf\{t\ge0:X(t)\in U_{\textup{target}}\}.
\end{align}
Note that $U_{\textup{target}}$ may be a disjoint union of multiple sets (i.e. the ``target'' $U_{\textup{target}}$ may consist of ``multiple targets''). 
If there are $N\ge1$ independent and identically distributed (iid) searchers with respective FPTs $\tau_1,\dots,\tau_N$, then the fastest FPT is
\begin{align}\label{eq:TN}
T_{N}
:=\min\{\tau_{1},\dots,\tau_{N}\},
\end{align}
where $\tau_{1},\dots,\tau_{N}$ are $N$ iid realizations of $\tau$ in \eqref{eq:tau0}. More generally, if we set $T_{1,N}:=T_N$, then the $k$th fastest FPT is
\begin{align}\label{eq:Tkn}
T_{k,N}
:=\min\big\{\{\tau_{1},\dots,\tau_{N}\}\backslash\cup_{j=1}^{k-1}\{T_{j,N}\}\big\},\quad k\in\{1,\dots,N\}.
\end{align}

In this chapter, we review recent results on the distribution and statistics of $T_N$ and $T_{k,N}$ for $1\le k\ll N$. We consider searchers which move by diffusion in section~\ref{sec:diffusion}, superdiffusion in section~\ref{sec:superdiffusion}, subdiffusion in section~\ref{sec:subdiffusion}, and jumps on a discrete network in section~\ref{sec:networks}. We conclude in section~\ref{sec:discussion} by discussing some closely related problems.

\section{Diffusion}\label{sec:diffusion}

We now study fastest FPTs for diffusive searchers. We start with a simple one-dimensional example.

\subsection{An introductory example: diffusion in one dimension}\label{sec:1d}

Suppose $X=\{X(t)\}_{t\ge0}$ is a one-dimensional pure diffusion process with diffusivity $D>0$. Let $\tau$ denote the FPT of $X$ to the origin (i.e.\ $\tau$ is in \eqref{eq:tau0} with $U_{\textup{target}}=0$). Assuming $X(0)=L>0$, the distribution of $\tau$ is given in terms of the error function,
\begin{align}\label{eq:erf}
    \P(\tau>t)
    =\textup{erf}\Big(\frac{L}{\sqrt{4Dt}}\Big).
\end{align}

Now, the mean of any nonnegative random variable $Z\ge0$ is given by the integral of its survival probability, $\E[Z]=\int_0^\infty \P(Z>z)\,\dd z$. Hence, the mean time for a single searcher to reach the origin is infinite,
\begin{align*}
    \E[\tau]
    =\int_0^\infty \P(\tau>t)\,\dd t
    =\infty,
\end{align*}
due to the slow large-time decay of $\P(\tau>t)$ in \eqref{eq:erf},
\begin{align}\label{eq:slow}
\P(\tau>t)
\sim (L/\sqrt{\pi D })t^{1/2}\quad\text{as }t\to\infty,
\end{align}
where throughout this chapter, $f\sim g$ denotes $f/g\to1$. 
Further, 
\begin{align}\label{eq:ET1N}
    \E[T_{N}]
    =\int_0^\infty [\P(\tau>t)]^N\,\dd t,
\end{align}
since $\P(T_N>t)=\P(\min\{\tau_1,\dots,\tau_N\}>t)=[\P(\tau>t)]^N$, where we have used that $\tau_1,\dots,\tau_N$ are iid. 
Therefore, \eqref{eq:slow} and \eqref{eq:ET1N} imply that $\E[T_{N}]$ is infinite if $N=2$, but finite for $N\ge3$ \cite{lindenberg1980}.

To obtain the large $N$ asymptotics of $\E[T_{N}]$, we combine \eqref{eq:erf} and \eqref{eq:ET1N} to yield
\begin{align}\label{eq:1dintegral}
    \E[T_{N}]
    =\int_0^\infty \Big[\textup{erf}\Big(\frac{L}{\sqrt{4Dt}}\Big)\Big]^N\,\dd t
    \sim \int_0^\eps \Big[\textup{erf}\Big(\frac{L}{\sqrt{4Dt}}\Big)\Big]^N\,\dd t\quad\text{as }N\to\infty
\end{align}
for any $\eps>0$, since the part of integral from $t=\eps$ to $t=\infty$ vanishes exponentially fast as $N\to\infty$. Using the large $z$ asymptotics of $\textup{erf}(z)$, it follows that \cite{meerson2015}
\begin{align}\label{eq:1d}
    \E[T_{N}]
    \sim\frac{L^2}{4D\ln N}\quad\text{as }N\to\infty.
\end{align}
In the next two subsections, we review how the result in \eqref{eq:1d} for this simple example extends to much more general scenarios.

\subsection{Higher spatial dimensions}

The basic result in \eqref{eq:1d} has been extended to bounded 2-dimensional and 3-dimensional spatial domains using probabilistic methods to prove that the fastest searchers take a direct path to the target \cite{lawley2020esp1}. This result is easiest to describe in the case that the spatial domain is a cylinder. Specifically, let the domain $U\subset\R^{3}$ be a cylinder of radius $r>0$ and height $h>0$,
\begin{align*}
U:=\big\{(x,y,z)\in\R^{3}:x^{2}+y^{2}<r^{2},\,z\in(0,h)\big\}.
\end{align*}
Suppose the target is a disk of radius $a\in(0,r)$ at the bottom of the cylinder,
\begin{align*}
U_\target:=\big\{(x,y,0)\in\R^{3}:x^{2}+y^{2}<a^{2}\big\}.
\end{align*}
Suppose that $N\ge1$ searchers are initially placed at $(0,0,{{L}})$ with $L\in(0,h)$ and then diffuse in $U$ with diffusivity $D>0$ and reflecting boundary conditions. 

Let $Z_{n}(t)\in[0,h]$ and $R_{n}(t)\in[0,r]$ denote the height and radial position of the $n$-th searcher at time $t\ge0$. The first time that the $n$-th searcher hits the target is
\begin{align*}
\tau_{n}^{\threed}:=\inf\{t>0:Z_{n}(t)=0,\;R_{n}(t)<a\},\quad n\in\{1,\dots,N\}.
\end{align*}
The first time that the $n$-th searcher hits the bottom of the cylinder (regardless of the radial position) is
\begin{align*}
\tau_{n}^{\oned}:=\inf\{t>0:Z_{n}(t)=0\},\quad n\in\{1,\dots,N\}.
\end{align*}
Hence, the first time that any searcher hits the target and the first time that any searcher hits the bottom of the cylinder are respectively given by
\begin{align*}
T_N^{\threed}
&:=\min_{n}\{\tau_{n}^{\threed}\},\quad
T_N^{\oned}
:=\min_{n}\{\tau_{n}^{\oned}\}.
\end{align*}
The next result shows that the moments of $T_N^{\threed}$ and $T_N^{\oned}$ become identical as $N$ grows.

\begin{theorem}[From reference~\cite{lawley2020esp1}]\label{thm:cylinder}
For any moment $m\ge1$, we have that
\begin{align}\label{eq:3d1d}
\E[(T_N^{\threed})^{m}]
\sim\E[(T_N^{\oned})^{m}]
\sim\Big(\frac{{{L}}^{2}}{4D\ln N}\Big)^{m}\quad\text{as }N\to\infty.
\end{align}
\end{theorem}

We make four comments on Theorem~\ref{thm:cylinder}. First, the proof of Theorem~\ref{thm:cylinder} relies on proving that the path of the first searcher to reach the target is almost a straight line from the initial position to the target. In particular, the fastest searcher out of $N\gg1$ searchers never leaves a tube of radius $a>0$ connecting the starting location to the target, and therefore $T_N^{\threed}=T_N^{\oned}$, and the large $N$ behavior of the moments of $T_N^{\oned}$ are already known since $T_N^{\oned}$ concerns diffusion in one space dimension \cite{yuste2001}.

Second, Theorem~\ref{thm:cylinder} holds for any fixed target size $a>0$.

Third, the cylinder is finite (i.e.\ $r<\infty$ and $h<\infty$). In fact, if $r=h=\infty$, then \eqref{eq:3d1d} cannot hold since $\E[T_N^{\threed}]=\infty$ for all $N\ge1$. To prove this, note that each searcher has a strictly positive probability of never reaching the target (i.e.\ $\P(\tau^\threed=\infty)>0$) since three-dimensional Brownian motion is transient.
Therefore, for any $N\ge1$, there is a strictly positive probability that no searcher hits the target,
\begin{align*}
\P(T_N^{\threed}=\infty)
=[\P(\tau^{\threed}=\infty)]^{N}>0,
\end{align*}
and therefore $\E[T_N^{\threed}]=\infty$ for all $N\ge1$. This is distinct from the phenomenon where the mean FPT of a single searcher is infinite while the MFPT of the fastest searcher is finite (see section~\ref{sec:1d}).

Finally, while Theorem~\ref{thm:cylinder} concerns a specialized spatial domain (i.e.\ a cylinder), the argument can be extended to prove that \eqref{eq:3d1d} holds in much more general $2$ and $3$-dimensional domains by considering a thin tube of length $L$ connecting the starting location to the target (see Theorems~5 and 10 in \cite{lawley2020esp1} for a precise statement). However, three restrictive assumptions of this argument are that (i) the searchers all start at a single point, (ii) the domain contains a straight-line path from this single starting location to the target, and (iii) the searchers move by pure diffusion. The next section shows that \eqref{eq:3d1d} still holds without these three assumptions.

\subsection{A universal moment formula}\label{sec:uni}

Looking back to the one-dimensional problem in section~\ref{sec:1d}, the formula for $\P(\tau>t)$ in \eqref{eq:erf} yields the following short-time behavior of $\P(\tau\le t)$ on a logarithmic scale,
\begin{align}\label{eq:log}
    \lim_{t\to0+}t\ln\P(\tau\le t)
    =-C<0,\quad\text{where }C:=\frac{L^2}{4D}>0.
\end{align}
Due to Varadhan's formula from large deviation theory \cite{varadhan1967, varadhan1967b}, the short-time behavior of the distribution of $\tau$ in \eqref{eq:log} holds in much more general scenarios \cite{lawley2020uni}, including $d$-dimensional diffusion processes (i) with general space-dependent diffusivities and drift fields, (ii) on Riemannian manifolds, (iii) in the presence of reflecting obstacles, and (iv) with partially absorbing targets. In these scenarios, $D>0$ is a characteristic diffusivity and $L>0$ is a certain geodesic distance between the searcher starting location(s) (possibly a distribution on a set of starting locations) and the target that (i) avoids any obstacles, (ii) includes any spatial variation or anisotropy in diffusivity, and (iii) incorporates any geometry in the case of diffusion on a curved manifold \cite{lawley2020uni}. Further, $L$ is unaffected by deterministic forces on the diffusive searchers (i.e.\ a drift) or a partially absorbing target \cite{lawley2020uni}.

Since the large $N$ behavior of $\E[T_{N}]$ depends only on the short-time behavior of $\P(\tau\le t)$ (see \eqref{eq:1dintegral}), the general formula in \eqref{eq:log} for the short-time behavior of $\P(\tau\le t)$ for diffusive search suggests that the large $N$ asymptotics of $\E[T_{N}]$ in \eqref{eq:1d} should extend to the aforementioned general scenarios in which \eqref{eq:log} holds. The following theorem shows that this is indeed the case, and in fact the result extends to the $m$th moment of the $k$th fastest FPT $T_{k,N}$ for $1\le k\ll N$.

\begin{theorem}[From reference~\cite{lawley2020uni}]\label{thm:uni}
Let $\{\tau_{n}\}_{n=1}^{\infty}$ be a sequence of iid nonnegative random variables. Assume that
\begin{align}\label{eq:conditiona}
\int_{0}^{\infty}\big[\P(\tau>t)\big]^{N}\,\dd t<\infty\quad\text{for some $N\ge1$}, 
\end{align}
and assume that there exists a constant $C>0$ so that
\begin{align}\label{eq:conditionb}
\lim_{t\to0+}t\ln\P(\tau\le t)=-C<0.
\end{align}
Then for any $m\ge1$ and $k\ge1$, the $m$th moment of $T_{k,N}$ in \eqref{eq:Tkn} satisfies
\begin{align}\label{eq:res}
\E[(T_{k,N})^{m}]
\sim\Big(\frac{C}{\ln N}\Big)^{m}\quad\text{as }N\to\infty.
\end{align}
\end{theorem}

Theorem~\ref{thm:uni} is a general result that applies to order statistics of any random variable satisfying \eqref{eq:conditiona}-\eqref{eq:conditionb}. Therefore, Theorem~\ref{thm:uni} and the universal behavior in \eqref{eq:log} for diffusive search shows that \eqref{eq:res} holds with $C=L^2/(4D)$ and the various details in the problem (spatial dimension, drift, domain size, target size, searcher number $k$, etc.)\ are irrelevant to the leading order statistics of $T_{k,N}$ as $N\to\infty$.

However, the convergence rate of \eqref{eq:res} is generally quite slow. Hence, it is important to identify how these various details affect the fastest FPT statistics at higher order. The next subsection computes these corrections using extreme value theory.

\subsection{Extreme value theory approach}

Extreme value theory is a branch of probability theory and statistics dealing with extreme events in the tails of probability distributions. The theory dates back nearly a century to Fisher, Tippett, Gnedenko \cite{fisher1928, colesbook} and is concerned with determining the distribution of the minima (or maxima) of a large sequence of iid random variables in terms of the short-time distribution of a single random variable. We now apply this theory to FPTs of diffusion.

Looking back again to the simple example in section~\ref{sec:1d}, the formula in \eqref{eq:erf} yields the following short-time behavior of $\P(\tau\le t)$ on a linear scale (i.e.\ more information than the logarithmic scale in \eqref{eq:log}),
\begin{align}\label{eq:short}
    \P(\tau\le t)
    \sim At^pe^{-C/t}\quad\text{as }t\to0+,
\end{align}
where $C=L^2/(4D)>0$, $A=\sqrt{4D/(L^2\pi)}$, and $p=1/2$. It turns out that this short-time behavior of the survival probability of diffusive search extends to much more general scenarios, where $C=L^2/(4D)$ is the diffusion timescale in \eqref{eq:log} and $A>0$ and $p\in\R$ depend on the details of the problem (spatial dimension, drift, target reactivity, etc.). The following theorem gives the distribution and statistics of $T_N$ using the information in \eqref{eq:short}.

\begin{theorem}[From reference~\cite{lawley2020dist}]\label{thm:dist}
Let $\{\tau_{n}\}_{n\ge1}$ be iid and assume that there exists constants $C>0$, $A>0$, and $p\in\R$ so that \eqref{eq:short} holds. Then $T_N:=\min\{\tau_1,\dots,\tau_N\}$ satisfies
\begin{align*}
\frac{T_{N}-b_{N}}{a_{N}}
\to_{\textup{d}}
X=_{\textup{d}}\textup{Gumbel}(0,1)\quad\text{as }N\to\infty,
\end{align*}
where 
\begin{align}\label{ababuf}
\begin{split}
a_{N}
&=\frac{C}{(\ln N)^{2}},\quad
b_{N}
=\frac{C}{\ln N}\Big(1+\frac{p\ln(\ln(N))}{\ln N}
-\frac{\ln(AC^{p})}{\ln N}\Big).
\end{split}
\end{align}
If we assume further that $\E[T_{N}]<\infty$ for some $N\ge1$, then for each $m\in(0,\infty)$,
\begin{align*}
\E\left[\left(\frac{T_{N}-b_{N}}{a_{N}}\right)^{m}\right]
\to\E[X^{m}]\quad\text{as }N\to\infty,\quad\text{where }X=_{\textup{d}}\textup{Gumbel}(0,1).
\end{align*}
\end{theorem}

We note that Theorem~\ref{thm:dist} can be generalized to describe $T_{k,N}$ for $1\le k\ll N$ (see Theorems~4 and 5 in \cite{lawley2020dist}). 

In Theorem~\ref{thm:dist} and throughout this chapter, $\to_\dd$ denotes convergence in distribution \cite{billingsley2013} and $=_\dd$ denotes equality in distribution. Further, $\textup{Gumbel}(0,1)$ denotes a random variable with a Gumbel distribution with location parameter $b=0$ and scale parameter $a=0$. Generally, a random variable $X$ has a Gumbel distribution with location parameter $b\in\R$ and scale parameter $a>0$ if\footnote{A Gumbel distribution is sometimes defined differently by saying that $-X$ has a Gumbel distribution with shape $-b$ and scale $a$ if \eqref{xgumbel} holds.}
\begin{align}\label{xgumbel}
\P(X>x)
=\exp\Big[-\exp\Big(\frac{x-b}{a}\Big)\Big],\quad\text{for all } x\in\R.
\end{align}
Note that if $X=_\dd \textup{Gumbel}(b,a)$, then $\E[X]=b-\gamma a$ where $\gamma\approx0.5772$ is the Euler-Mascheroni constant, and $\textup{Variance}(X)=\frac{\pi^{2}}{6}a^{2}$. Theorem~\ref{thm:dist} thus yields higher order estimates of statistics of $T_{N}$. In particular, 
\begin{align*}
\E[T_{N}]
&=b_{N}-\gamma a_{N}+o(a_{N}),\\
\textup{Variance}(T_{N})
&=\frac{\pi^{2}}{6}a_{N}^{2}+o(a_{N}^{2}).
\end{align*}
Roughly speaking, Theorem~\ref{thm:dist} implies that $T_{N}$ is approximately Gumbel with shape $b_{N}$ and scale $a_{N}$.

To illustrate Theorem~\ref{thm:dist}, consider one-dimensional diffusive search as in section~\ref{sec:1d}, but now suppose there is a constant drift $V>0$ pushing the searchers toward the target at $x=0$. That is, the position of a searcher evolves according to the stochastic differential equation,
\begin{align}\label{drift}
\dd X
=-V\,\dd t+\sqrt{2D}\,\dd W,\quad X(0)=L>0,
\end{align}
where $W$ is a standard Brownian motion. If $\tau$ is the FPT to the origin (i.e.\ \eqref{eq:tau0} with $U_\target=0$), then the survival probability is
\begin{align}\label{eq:SV}
\P(\tau>t)
=\frac{1}{2}\bigg[1+\textup{erf}\Big(\frac{L-Vt}{\sqrt{4Dt}}\Big)
-e^{\frac{VL}{D}}\textup{erfc}\Big(\frac{L+Vt}{\sqrt{4Dt}}\Big)\bigg],\quad t>0.
\end{align}
Using \eqref{eq:SV}, one can check that \eqref{eq:short} holds with $C=L^2/(4D)$, $A=\sqrt{1/(C\pi)}e^{LV/(2D)}$, and $p=1/2$. Hence, Theorem~\ref{thm:dist} implies that as $N\to\infty$,
\begin{align}\label{eq:V}
    \E[T_{N}]
    =\frac{L^2}{4D\ln N}\Big[1+\frac{\ln(\ln(N))}{2\ln N}-\Big(\frac{LV/D-\ln\pi+2\gamma}{\ln N}\Big)+o((\ln N)^{-1})\Big].
\end{align}
This result shows that increasing $V$ (i.e.\ increasing the drift toward the target) decreases $\E[T_{N}]$. This is to be expected, but \eqref{eq:V} shows that the effect of increasing $V$ only affects $\E[T_{N}]$ at third order as $N\to\infty$. 

We note that the analysis above of how $T_N$ vanishes is all in the limit $N\to\infty$ for a fixed drift, fixed target size, etc. However, it is clear that $T_{N}$ must diverge if we fix $N\ge1$ (even a large value of $N$) and take, for example, the size of the target to zero (similarly, $T_{N}$ diverges if we take the target reactivity to zero or take a repelling drift pushing the searchers away from the target to infinity). For an analysis of the ``competing limits'' of $N\to\infty$ versus, say, a vanishing target (or an unreactive target or a large repulsive drift), we refer the reader to Reference~\cite{madrid2020comp}.

\subsection{Uniform initial conditions}

The analysis above all assumes that the diffusive searchers cannot start arbitrarily close to the target. That is, even in the case where the searchers may have a continuum of possible starting locations (as in section~\ref{sec:uni}), the support of this set of starting locations is a strictly positive distance away from the closest part of the target. For example, the one-dimensional example in section~\ref{sec:1d} could be modified so that the initial distribution of $X$ is uniform on the interval $[a,b]$ as long as $0<a<b<\infty$. In this case, the moment formula in \eqref{eq:res} holds with $C=L^2/(4D)$ and $L=a>0$.

How does the fastest FPT change if searchers can start arbitrarily close to the target? For example, suppose $X$ is initially uniformly distributed on the interval $[0,l]$ with targets at both $x=0$ and $x=l>0$ (i.e.\ $\tau$ is in \eqref{eq:tau0} with $U_\target=\{0,l\}$). In this case, it was shown back in the first work on fastest FPTs of diffusion that \cite{weiss1983}
\begin{align}\label{eq:Nsquared}
    \E[T_N]
    \sim\frac{\pi l^2}{8D}\frac{1}{N^2}\quad\text{as }N\to\infty,
\end{align}
which is of course much faster than the $1/\ln N$ decay seen in the sections above for the case that searchers cannot start arbitrarily close to the target.

The much faster decay of $\E[T_N]$ for this problem stems from the following short-time behavior of the distribution of $\tau$ \cite{madrid2020comp},
\begin{align}\label{eq:shortuniform}
    \P(\tau\le t)
    \sim At^p\quad\text{as }t\to0+,
\end{align}
where $p=1/2$ and $A=\sqrt{4^2D/(l^2\pi)}$, which is much slower than the short-time decay of $\P(\tau\le t)$ in \eqref{eq:log}. In fact, rather than just the mean in \eqref{eq:Nsquared}, it follows from \eqref{eq:shortuniform} and Theorem~\ref{thm:evtpower} below that $(AN)^{1/p}T_N$ converges in distribution to a Weibull random variable with unit scale and shape $p$ as $N\to\infty$ 

The decay of $\E[T_N]$ for this example slows from $1/N^2$ in \eqref{eq:Nsquared} to $1/N$ if the targets at $x=0$ and $x=l>0$ are partially absorbing with reactivity $\kappa\in(0,\infty)$. Specifically, it was shown in \cite{madrid2020comp, grebenkov2020} that
\begin{align}\label{eq:1overN}
    \E[T_N]
    \sim\frac{l}{2\kappa}\frac{1}{N} \quad\text{as }N\to\infty.
\end{align}
The behavior in \eqref{eq:1overN} follows from the fact that the short-time distribution of $\tau$ satisfies \eqref{eq:shortuniform} with $p=1$ and $A=2\kappa/l$ \cite{madrid2020comp}. In fact, rather than just the mean in \eqref{eq:1overN}, it follows from \eqref{eq:shortuniform} and Theorem~\ref{thm:evtpower} below that $(2\kappa/l)NT_N$ converges in distribution to a unit rate exponential random variable as $N\to\infty$. 

\section{Superdiffusion}\label{sec:superdiffusion}

Section~\ref{sec:diffusion} concerns fastest FPTs of normal diffusion processes. Normal diffusion is marked by a squared displacement that grows linearly in time. However, rather than normal diffusion, an anomalous form of diffusion called superdiffusion has been observed in a variety of physical and biological systems \cite{metzler2004}. Superdiffusion is marked by a squared displacement that grows superlinearly in time. In this section, we investigate fastest FPTs for a model of superdiffusion called a L{\'e}vy flight \cite{dubkov2008}.

\subsection{L{\'e}vy flights}\label{sec:levyflights}

A L{\'e}vy flight with stability index $\alpha\in(0,2)$ can be obtained from the continuous-time random walk model \cite{montroll1965, metzler2004} assuming that the waiting time between jumps of the walker has a finite mean $t_0\in(0,\infty)$ and the probability density of the jump distance has the following slow power law decay,
\begin{align}\label{pl}
f(y)
\sim
 \frac{(l_0)^{\alpha}}{y^{1+\alpha}}\quad\text{as $y\to\infty$ for some lengthscale $l_0>0$}.
\end{align}
The probability density ${{p}}(x,t)$ for the position of a L{\'e}vy flight satisfies the following space-fractional Fokker-Planck equation \cite{meerschaert2011},
\begin{align}\label{ffpe}
\frac{\partial}{\partial t}p
=-K(-\Delta)^{\alpha/2}p,
\end{align}
where $K=(l_0)^{\alpha}/t_0>0$ is the generalized diffusivity and $(-\Delta)^{\alpha/2}$ denotes the fractional Laplacian \cite{lischke2020}. A L{\'e}vy flight $X$ in $\R^d$ can also be defined as the following random time change (i.e.\ a subordination) of a Brownian motion \cite{meerschaert2011}, 
\begin{align}\label{eq:supertimechange}
    X(t)
    :=B(S(t))+X(0)\quad t\ge0,
\end{align}
where $B=\{B(s)\}_{s\ge0}$ is a $d$-dimensional Brownian motion with unit diffusivity (i.e.\ $\E\|{{B}}(s)\|^{2}=2ds$ for all $s\ge0$) and $S=\{S(t)\}_{t\ge0}$ is an independent, $(\alpha/2)$-stable subordinator with Laplace exponent $\Phi(\beta)=K\beta^{\alpha/2}$ (i.e.\ $\E[e^{-\beta S(t)}]=e^{-t\Phi(\beta)}$ for all $t\ge0$, $\beta\ge0$). 

\subsection{Fastest FHTs of L{\'e}vy flights}

Define the first hitting time (FHT) $\tau$ of the L{\'e}vy flight $X$ in \eqref{eq:supertimechange} to some target $U_\target\subset\R^d$ as in \eqref{eq:tau0}. We use the term FHT rather than FPT since these two concepts can be distinct for L{\'e}vy flights due to their discontinuous sample paths \cite{koren2007, koren2007first, palyulin2019, wardak2020}. As above, determining the statistics and distribution of $T_{k,N}$ for $1\le k\ll N$ in \eqref{eq:Tkn} requires determining the short-time distribution of $\tau$.

Assuming that $X(0)$ cannot lie in the target $U_\target$, it was proven in \cite{lawley2023super} that $\tau$ has the universal short-time distribution,
\begin{align}\label{eq:shortsuper}
\P(\tau\le t)
\sim \rho t\quad\text{as }t\to0+,
\end{align}
where $\rho\in(0,\infty)$ is the rate,
\begin{align}\label{rho0}
\rho
:=\int_{0}^{\infty}\P({{B}}(s)+X(0)\in U_\target)\frac{\alpha/2}{\Gamma(1-\alpha/2)}\frac{{{K}}}{s^{1+\alpha/2}}\,\dd s.
\end{align}
If $X(0)=0$, then the Gaussianity of ${{B}}(s)$ means that $\rho$ can be written as
\begin{align*}
\rho
=\int_0^\infty \frac{1}{(4\pi s)^{d/2}} \int_{U_\target}\exp\Big(\frac{-\|x\|^{2}}{4s}\Big)\,\dd x \,\frac{\alpha/2}{\Gamma(1-\alpha/2)}\frac{{{K}}}{s^{1+\alpha/2}}\,\dd s.
\end{align*}

The following theorem yields the distribution and statistics of $T_{N}$ in \eqref{eq:TN} from the short-time behavior in \eqref{eq:shortsuper}. Analogous results hold for $T_{k,N}$ in \eqref{eq:Tkn} with $1\le k\ll N$ (see Theorems~5 and 6 in \cite{madrid2020comp}).

\begin{theorem}[From reference~\cite{madrid2020comp}]\label{thm:evtpower}
Let $\{\tau_{n}\}_{n\ge1}$ be an iid sequence of random variables and assume that for some ${{A}}>0$ and $p>0$, we have that
\begin{align}\label{short}
\P(\tau_{n}\le t)
&\sim{{A}} t^{p}\quad\text{as }t\to0+.
\end{align}
Then, the following rescaling of $T_{N}:=\min\{\tau_{1},\dots,\tau_{N}\}$ converges in distribution,
\begin{align*}
(AN)^{1/p}T_{N}
\to_{\dd}
\textup{Weibull}(1,p)\quad\text{as }N\to\infty.
\end{align*}
    If we assume further that $\E[T_{N}]<\infty$ for some $N\ge1$, 
then for each $m\in(0,\infty)$,
\begin{align*}
\E[(T_{N})^{m}]
&\sim \frac{\Gamma(1+m/p)}{(AN)^{m/p}}\quad\text{as }N\to\infty.
\end{align*}
\end{theorem}

Throughout this chapter, $X=_\dd\textup{Weibull}(t,p)$ means that $X$ is a Weibull random variable with scale $t>0$ and shape $p>0$, which means
\begin{align}\label{xweibull}
\P(X>x)
=\exp(-(x/t)^{p}),\quad x\ge0.
\end{align}
Note that $\textup{Weibull}(1,p)$ is exponential with unit rate if $p=1$. Therefore, applying Theorem~\ref{thm:evtpower} to \eqref{eq:shortsuper} implies that $(\rho N)T_{N}$ converges in distribution to a unit rate exponential random variable as $N$ grows \cite{lawley2023super}. 
Hence, $T_{N}$ is approximately exponentially distributed with rate $\rho N$. Furthermore, if $\E[T_{N}]<\infty$ for some $N\ge1$, then Theorem~\ref{thm:evtpower} yields all the moments of $T_{N}$ for large $N$. In particular,
\begin{align}\label{eq:decay}
\E[T_{N}]
&\sim\sqrt{\text{Variance}[T_{N}]}
\sim\frac{1}{\rho N}\quad\text{as }N\to\infty.
\end{align}
Analogous results hold for (i) $T_{k,N}$ in \eqref{eq:Tkn} for any $1\le k\ll N$ and (ii) the case that $S$ is any nondeterministic L{\'e}vy subordinator \cite{lawley2023super}.

We now contrast \eqref{eq:decay} with the case of normal diffusion. First, the $1/N$ decay in \eqref{eq:decay} is much faster than the $1/\ln N$ decay for normal diffusion processes in section~\ref{sec:diffusion}.  Further, fastest FPTs of diffusion depend on the shortest path to the target since the fastest diffusive searchers follow this geodesic \cite{lawley2020uni}. Indeed, fastest FPTs of diffusion are unaffected by changes to the problem outside of this path, such as altering the domain or target size or even the spatial dimension. In contrast, the expression for the rate $\rho$ in \eqref{rho0} implies that fastest FHTs of L{\'e}vy flights depend on these global properties of the problem. Indeed, the fastest L{\'e}vy flights do not take a direct path to the closest part of the target \cite{lawley2023super}.

\section{Subdiffusion}\label{sec:subdiffusion}

Another form of anomalous diffusion seen in a variety of physical and biological systems is subdiffusion \cite{oliveira2019, klafter2005, hofling2013, barkai2012}. Subdiffusion is marked by a mean-squared displacement that grows according to the following sublinear power law,
\begin{align*}
\E\|X(t)-X(0)\|^{2}
\,\propto\, t^{\gamma},\quad\gamma\in(0,1).
\end{align*}

\subsection{Subordination}

One common way to model subdiffusion is via a time-fractional Fokker-Planck equation \cite{metzler1999}, which derives from the continuous-time random walk model with infinite mean waiting times between jumps \cite{montroll1965, metzler2000}. Such a time-fractional Fokker-Planck equation describes the distribution of a subdiffusive process $\{X(t)\}_{t\ge0}$ defined by subordinating a normal diffusive process $\{Y(s)\}_{s\ge0}$ \cite{lawley2020subpre}. Precisely, if $\{Y(s)\}_{s\ge0}$ is a normal diffusion process satisfying an It\^{o} stochastic differential equation, then \begin{align}\label{eq:XYS}
X(t)
:=Y(S(t)),\quad t\ge0,
\end{align}
where $S=\{S(t)\}_{t\ge0}$ is an independent, inverse $\gamma$-stable subordinator. We emphasize that $S$ is an \textit{inverse} stable subordinator in this section (in contrast with section~\ref{sec:superdiffusion}).

The definition of $X$ in \eqref{eq:XYS} implies that if $\tau$ and $\sigma$ are respective FPTs of $X$ and $Y$ to some target, then the distribution of $\tau$ is
\begin{align}\label{eq:rep}
\P(\tau\le t)
=\E[F_\sigma(S(t))].
\end{align}
where $F_\sigma(s):=\P(\sigma\le s)$ is the cumulative distribution function of the FPT $\sigma$ of the normal diffusion process $Y$. 
It follows immediately from \eqref{eq:rep} that
\begin{align}\label{eq:reprep}
\P(\tau\le t)
=\int_{0}^{\infty}\P(\sigma\le s)\frac{t}{\gamma s^{1+1/\gamma}}l_{\gamma}\Big(\frac{t}{s^{1/\gamma}}\Big)\,\dd s,
\end{align}
since the probability density that $S(t)=s$ is 
\begin{align}\label{lspdf}
\frac{\dd}{\dd s}\P(S(t)\le s)
=\frac{t}{\gamma s^{1+1/\gamma}}l_{\gamma}\Big(\frac{t}{s^{1/\gamma}}\Big),
\end{align}
where $l_{\gamma}(z)$ is defined by its Laplace transform,
\begin{align*}
\int_{0}^{\infty}e^{-rz}l_{\gamma}(z)\,\dd z
=e^{-r^{\gamma}},\quad \gamma\in(0,1),\; r\ge0.
\end{align*}

\subsection{Fastest FPTs of subdiffusion}

By using \eqref{eq:reprep}, the small $s$ behavior of $\P(\sigma\le s)$, and the small $z$ behavior of $l_\gamma(z)$, we can develop a theory of fastest FPTs of subdiffusion that is analogous to the theory for normal diffusion in section~\ref{sec:diffusion}. In particular, since $\sigma$ is a FPT of a normal diffusion process, then we have that under very general conditions (see section~\ref{sec:uni}),
\begin{align*}
    \lim_{s\to0+}s\ln\P(\sigma\le s)
    =-L^2/(4D)<0,
\end{align*}
where $D$ and $L$ are as in section~\ref{sec:uni} for the normal diffusion process $Y$ (note that $Y=Y(s)$ is indexed by ``internal time'' $s\ge0$, which is not physical time, but rather has dimension $(\text{time})^{\gamma}$ and thus $D$ has dimension $(\text{length})^2(\text{time})^{-\gamma}$). Then, using \eqref{eq:reprep} and the fact that \cite{schneider1986, barkai2001}
\begin{align}\label{logl}
\lim_{z\to0+}z^{\gamma/(1-\gamma)}\ln l(z)
=-(1-\gamma)\gamma^{\gamma/(1-\gamma)}<0,
\end{align}
we obtain the following universal behavior of FPTs of subdiffusion modeled by a time-fractional Fokker-Planck equation,
\begin{align}\label{eq:shortsub}
    \lim_{t\to0+}t^{\gamma/(2-\gamma)}\P(\tau\le t)
    =-C
    :=-(2-\gamma) \gamma ^{\frac{\gamma }{2-\gamma}} (L^2/(4D))^{\frac{1}{2-\gamma }}<0.
\end{align}

By a result analogous to Theorem~\ref{thm:uni} above (see Theorem~6 in \cite{lawley2020sub}), the short-time behavior on a logarithmic scale in \eqref{eq:shortsub} implies that the $m$th moment of $T_{k,N}$ in \eqref{eq:Tkn} for subdiffusion satisfies (assuming $\E[T_N]<\infty$ for some $N\ge1$),
\begin{align}\label{eq:submoment}
    \E[(T_{k,N})^{m}]
    \sim\bigg(\frac{t_{\gamma}}{(\ln N)^{2/\gamma-1}}\bigg)^{m}
    \quad\text{as }N\to\infty,
\end{align}
where $t_{\gamma}>0$ is the characteristic subdiffusive timescale,
\begin{align*}
    t_{\gamma}
    :=\Big(\gamma^{\gamma}(2-\gamma)^{2-\gamma}\frac{L^{2}}{4D}\Big)^{1/\gamma}>0,\quad \gamma\in(0,1].
\end{align*}
We can also generalize Theorem~\ref{thm:dist} above to determine higher order corrections to \eqref{eq:submoment} and the Gumbel probability distribution of $T_{k,N}$, but we omit these results for brevity (see Theorems 9, 10, and 11 in \cite{lawley2020sub}). Note also that we focused here on the case when searchers cannot start arbitrarily close to the target, but one can also determine the fastest FPTs of subdiffusive searchers which start uniformly in a bounded spatial domain (see the discussion section of \cite{lawley2020sub}).

\subsection{Extreme statistics: subdiffusion is faster than normal diffusion}

Comparing \eqref{eq:submoment} with \eqref{eq:res} in section~\ref{sec:diffusion} for normal diffusion yields the counterintuitive result that fastest FPTs of subdiffusion are faster than fastest FPTs of normal diffusion. Concretely, let $\sigma^{\textup{norm}}$ and $\tau^{\textup{sub}}$ denote FPTs to a target in a bounded domain for normal diffusive and subdiffusive search, respectively. We generally have
\begin{align*}
    \E[\sigma^{\textup{norm}}]
    <\E[\tau^{\textup{sub}}]=\infty,
\end{align*}
where $\E[\tau^{\textup{sub}}]=\infty$ owes to the slow decay of $\P(\tau^{\textup{sub}}>t)$ as $t\to\infty$ \cite{yuste2004}. However, if $\P(\sigma^{\textup{norm}}>s)$ decays at least as fast as a power law as $s\to\infty$, then \eqref{eq:reprep} implies that (see Theorem~8 in \cite{lawley2020sub})
\begin{align*}
    \E[T_N^{\textup{sub}}]<\infty\quad\text{for sufficiently large }N,
\end{align*}
where $T_N^{\textup{sub}}:=\min\{\tau_1^{\textup{sub}},\dots,\tau_N^{\textup{sub}}\}$ and $\{\tau_n^{\textup{sub}}\}_n$ are iid. Further, \eqref{eq:submoment} and \eqref{eq:res} imply
\begin{align*}
    \E[T_N^{\textup{sub}}]
    \ll\E[\Sigma_N^{\textup{norm}}]\quad\text{ for sufficiently large }N,
\end{align*}
where $\Sigma_N^{\textup{norm}}:=\min\{\sigma_1^{\textup{norm}},\dots,\sigma_N^{\textup{norm}}\}$ and $\{\sigma_n^{\textup{norm}}\}_n$ are iid.

\section{Random walks on networks}\label{sec:networks}

In sections~\ref{sec:diffusion}-\ref{sec:subdiffusion}, we considered search processes on a continuous-state space. We now follow \cite{lawley2020networks} and consider searchers which move on a ``network'' of discrete states. 
 
\subsection{Random walk setup}

Each searcher moves on the network according to a continuous-time Markov chain. Hence, the waiting time between jumps is exponentially distributed, but this assumption on the waiting time distribution can be relaxed (see section III in \cite{lawley2020networks}).

Concretely, let $X=\{X(t)\}_{t\ge0}$ be a continuous-time Markov chain on a finite or countably infinite state space $I$. The process $X$ is a single searcher and its dynamics are described by its infinitesimal generator matrix $Q=\{q(i,j)\}_{i,j\in I}$ \cite{norris1998}. The off-diagonal entries of $Q$ (i.e.\ $q(i,j)\ge0$ for $i\neq j$) are nonnegative and give the rate that $X$ jumps from $i\in I$ to $j\in I$. The diagonal entries of $Q$ are nonpositive (i.e.\ $q(i,i)\le0$ for all $i\in I$) and are chosen so that $Q$ has zero row sums. Assume that $\sup_{i\in I}|q(i,i)|<\infty$ so that $X$ cannot take infinitely many jumps in finite time.

\subsection{Single FPTs}

Let $\tau$ in \eqref{eq:tau0} be the FPT of $X$ to some target set $U_{\target}\subset I$. Let $\rho=\{\rho(i)\}_{i\in I}=\{\P(X(0)=i)\}_{i\in I}$ denote the initial distribution of $X$.
To avoid trivial cases, assume that $X$ cannot start on the target, which means that  $U_{\target}\cap\textup{supp}(\rho)=\varnothing$, where $\textup{supp}(\rho)\subset I$ denotes the support of $\rho$ (i.e.\ $\textup{supp}(\rho):=\{i\in I:\rho(i)>0\}$).

As in the sections above, determining the distribution of $T_N$ in \eqref{eq:TN} and $T_{k,N}$ in \eqref{eq:Tkn} requires determining the short-time distribution of $\tau$. Theorem~\ref{thm:rw} below shows that
\begin{align}\label{eq:basic}
\P(\tau\le t)
\sim\frac{\Lambda}{d!}t^{d}\quad\text{as }t\to0+,
\end{align}
where $d\ge1$ is the smallest number of jumps that $X$ must take to hit the target and $\Lambda>0$ is a sum of the products of the jump rates along the shortest path to the target. 

To make $d$ and $\Lambda$ in \eqref{eq:basic} precise, define a path $\PP$ of length $d\in\mathbb{Z}_{\ge0}$ from $i_{0}\in I$ to $i_{d}\in I$ to be a sequence of $d+1$ states in $I$,
\begin{align}\label{path}
\PP
=(\PP(0),\dots,\PP(d))
=(i_{0},i_{1},\dots,i_{d})\in I^{d+1},
\end{align}
so that
\begin{align}\label{explain}
q(\PP(k),\PP(k+1))
>0,\quad\text{for }k\in\{0,1,\dots,d-1\}.
\end{align}
The condition in \eqref{explain} means that $X$ has a strictly positive probability of following the path $\PP$. Assume that there is a path from the support of $\rho$ to the target (otherwise, $\P(\tau=\infty)=1$ and the problem is trivial).

For a path $\PP\in I^{d+1}$, let $\lambda(\PP)$ be the product of the rates along the path,
\begin{align}\label{lambda}
\lambda(\PP)
:=\prod_{i=0}^{d-1}q(\PP(i),\PP(i+1))>0.
\end{align}
Let $\dmin(I_{0},I_{1})\in\mathbb{Z}_{\ge0}$ denote the length of the shortest path from $I_{0}\subset I$ to $I_{1}\subset I$,
\begin{align*}
\dmin(I_{0},I_{1})
:=
\inf\{d:\PP\in I^{d+1},\PP(0)\in I_{0},\PP(d)\in I_{1}\}.
\end{align*}
That is, $\dmin(I_{0},I_{1})$ is the smallest number of jumps that $X$ must take to go from $I_{0}$ to $I_{1}$. 
Define the set of all paths from $I_{0}$ to $I_{1}$ with the length $\dmin(I_{0},I_{1})$,
\begin{align}\label{S}
\begin{split}
\mathcal{S}(I_{0},I_{1})
&:=\{\PP\in I^{d+1}:\PP(0)\in I_{0},\PP(d)\in I_{1},d=\dmin(I_{0},I_{1})\}.
\end{split}
\end{align}
Define
\begin{align}\label{Lambdathm}
\Lambda(\rho,I_{1})
:=\sum_{\PP\in\mathcal{S}(\textup{supp}(\rho),I_{1})}\rho(\PP(0))\lambda(\PP).
\end{align}
To explain $\Lambda(\rho,I_{1})$, suppose first that $\rho(i_{0})=1$ for some $i_{0}\in I$ (i.e.\ $\text{supp}(\rho)=i_{0}=X(0)$). If there is only one path with the minimum number of jumps $\dmin(i_{0},I_{1})$, then $\Lambda(\rho,I_{1})$ is the product of the jump rates along this path (i.e.\ $\lambda(\PP)$ in \eqref{lambda}). If there is more than one shortest path, then $\Lambda(\rho,I_{1})$ is the sum of the products of the jump rates along these paths. Finally, if $\rho$ is not concentrated at a single point, then $\Lambda(\rho,I_{1})$ merely sums the products of the jump rates along all the shortest paths, where the sum is weighted according to $\rho$. With these definitions in place, we can now give the short-time behavior of the distribution of $\tau$.

\begin{theorem}[From reference~\cite{lawley2020networks}]\label{thm:rw}
The short-time distribution of $\tau$ satisfies \eqref{eq:basic}, where
\begin{align*}
d
&=\dmin(\textup{supp}(\rho),U_{\target})\in\mathbb{Z}_{>0},\\
\Lambda
&=\Lambda(\rho,U_{\target})>0.
\end{align*}
\end{theorem}

\subsection{Fastest FPTs}

Having determined the short-time probability distribution of a single FPT in Theorem~\ref{thm:rw}, we can immediately determine the limiting distribution and statistics of fastest FPTs for random walks on networks. In particular, define
\begin{align*}
A
&=\frac{\Lambda}{d!}>0,
\end{align*}
where $d\ge1$ and $\Lambda>0$ are as in Theorem~\ref{thm:rw}. 
Theorem~\ref{thm:evtpower} implies that the following rescaling of $T_{N}$ converges in distribution to a Weibull random variable,
\begin{align}\label{eq:cd}
(AN)^{1/d}T_{N}
\to_\dd
\textup{Weibull}(1,d)\quad\text{as }N\to\infty.
\end{align}
Furthermore, if $\E[T_{N}]<\infty$ for some $N\ge1$, then for each moment $m\in(0,\infty)$, 
\begin{align}\label{eq:momentformula}
\E[(T_{N})^{m}]
&\sim \frac{\Gamma(1+m/d)}{(AN)^{m/d}}\quad\text{as }N\to\infty.
\end{align}

Compared to a single FPT $\tau$, fastest FPTs $T_N$ are faster and less variable since \eqref{eq:momentformula} implies that the mean and variance of $T_N$ vanish as $N\to\infty$. Furthermore, fastest FPTs are independent of much of the properties of the network. In particular, \eqref{eq:cd} implies that the large $N$ distribution of $T_{N}$ is fully determined by $N$, $\Lambda$, and $d$, which depend only on network features along the shortest path(s) from the initial distribution to the target. Similar conclusions hold for $T_{k,N}$ for $1\le k\ll N$ (see Theorems~2 and 4 in \cite{lawley2020networks}).

\section{Discussion}\label{sec:discussion}

In this chapter, we reviewed fastest FPTs out of many searchers for a variety of search processes. We found that such extreme FPTs tend to differ drastically from the behavior of a single FPT. We showed that the statistics and distribution of fastest FPTs are determined by the short-time distribution of a single FPT. Hence, details in the problem (spatial dimension, domain size and geometry, some aspects of searcher dynamics, etc.)\ can become irrelevant to fastest FPTs in the many searcher limit. These details become irrelevant because the fastest FPTs typically correspond to searchers which take a direct path to the target.

We now briefly mention a few closely related problems. The behavior of fastest searchers mimics so-called ``mortal'' or ``evanescent'' searchers \cite{abad2010, abad2012, abad2013, yuste2013, meerson2015b, meerson2015, grebenkov2017} which have been conditioned to reach the target before a fast inactivation time \cite{ma2020, lawley2021mortal, lawley2020networks}. Conditioning on not being inactivated effectively filters out searchers which do not take a direct path to the target. 

In addition to FPTs, cover times of multiple searchers have also been investigated \cite{chupeau2015, majumdar2016, dong2023, kim2023}. Cover times measure the speed of exhaustive search and are defined as the first time a searcher(s) comes within a specified ``detection radius'' of every point in the target region (often the entire spatial domain). For diffusive search, it was recently proven that the asymptotic $L^2/(4D\ln N)$ also holds for the time it takes $N\gg1$ iid searchers to collectively cover a target if $L$ is the shortest distance from the searcher starting locations to the \textit{farthest} part of the target \cite{kim2023}.

In the case of multiple targets, one can consider the probability that the fastest searcher hits a particular target first (the so-called splitting probability or hitting probability). In the limit of many diffusive searchers, the fastest searcher always hits the nearest target, regardless of target size, domain geometry, drift, etc.\ \cite{linn2022}. This phenomenon has been shown to offer a way for biological cells to detect the location of the source of a diffusing signal \cite{bernoff2023, lindsay2023} and also offers insight into the choices made by the first deciders in large groups of decision-makers \cite{karamched2020, stickler2023}. 

\textit{Slowest} FPTs have recently found application in ovarian aging and menopause timing \cite{lawley2023bor, lawley2023slow}. Women are born with $N\in[10^5,10^6]$ primordial follicles in their ovarian reserve, and this reserve then decays until only about $k=10^3$ follicles remain which is thought to trigger menopause \cite{richardson1987}. Further, there is a long history \cite{faddy1976, faddy1983, hirshfield1991, faddy1995, finch2000, johnson2022} of models that assume that follicles leave the reserve at iid random times $\tau_1,\dots,\tau_N$. Hence, the menopause age is $T_{N-k,N}$ (using the notation in \eqref{eq:Tkn}), which means that menopause age is determined by the  $k/N\approx0.2\%$ slowest follicles to leave the reserve. This analysis has been used to offer an explanation for the so-called ``wasteful'' oversupply of follicles \cite{lawley2023bor}.

Finally, we end by mentioning a possible problem with the analysis of fastest FPTs of diffusive searchers. In some scenarios, the vanishing mean fastest search time,
\begin{align}\label{eq:vanish}
    \E[T_N]
    \sim\frac{L^2}{4D\ln N}\quad\text{as }N\to\infty,
\end{align}
becomes unphysical. Specifically, if the searchers have a finite maximum speed $v>0$, then no searcher can reach a target that is distance $L>0$ from the starting location before time $L/v>0$. That is, we must have $T_N\ge L/v>0$, which contradicts \eqref{eq:vanish} for sufficiently large $N$. This contradiction stems from the well-known \cite{keller2004} problem of approximating finite speed random walks by diffusion due to the infinite speed of propagation of solutions to the diffusion equation \cite{kuske1997, joseph1989}. This discrepancy between finite speed random walks and infinite speed diffusion can often be ignored in many applications, since the discrepancy occurs in the tails of the distribution. However, these tails determine fastest FPTs. For a discussion and analysis of (i) when \eqref{eq:vanish} becomes unphysical and (ii) fastest FPTs of  run and tumble processes (i.e.\ finite speed random walks), we refer the reader to Reference~\cite{lawley2021pdmp}.

\bibliography{library.bib}
\bibliographystyle{unsrt}
\end{document}